\documentclass[runningheads]{llncs}
\usepackage[T1]{fontenc}
\usepackage{graphicx}
\usepackage{fancyhdr} 
\usepackage{siunitx}
\usepackage{amsmath,amssymb,amsfonts}
\usepackage{algorithmic}
\usepackage{graphicx}
\usepackage{textcomp}
\usepackage{graphicx}
\usepackage{amsmath,amssymb,amsfonts}
\usepackage{algorithmic}
\usepackage{graphicx}
\usepackage{textcomp}
\usepackage{wrapfig}
\usepackage{multicol}
\usepackage[T1]{fontenc}
\usepackage{booktabs} 
\usepackage{array}    
\usepackage{booktabs}
\usepackage{float}
\usepackage{siunitx} 
\usepackage{tabularx}
\usepackage{adjustbox} 
\usepackage[utf8]{inputenc}
\usepackage[backend=biber,style=ieee]{biblatex}
\usepackage{hyperref}
\hypersetup{
    colorlinks=true,       
    linkcolor=blue,        
    citecolor=blue,        
    urlcolor=blue          
}
\begin{document}
\title{MENGLAN:Multiscale Enhanced Nonparametric Gas
Analyzer with Lightweight Architecture and
Networks}
%
%

\author{Zhenke Duan, Jiqun Pan, Jiani Tu}
\authorrunning{Z. Duan et al.}
\institute{Zhongnan University of Economics and Law, Wuhan, China \and Wuhan SecureScape Technology Co., Ltd., Wuhan, China
\email{duanzhenke@sscapewh.com}
}



\maketitle              
\begin{abstract}
Accurate detection of ethylene concentrations in mixed
gases is crucial in chemical production for safety and health purposes. Traditional methods are hindered by high cost and complexity, limiting their practical application. This study proposes
MENGLAN, a Multiscale Enhanced Nonparametric Gas Analyzer that integrates a dual-stream structure, a Hybrid Multi-Head
Attention mechanism, and a Feature Reactivation Module to enable
real-time, lightweight, and high-precision ethylene concentration
prediction. Results show that MENGLAN achieves superior performance, reduced computational demand, and enhanced deployability compared to existing methods.

\end{abstract}
\section{INTRODUCTION}
Ethylene is a highly flammable and toxic gas. Accurate monitoring is essential to prevent accidents and protect health in industrial settings. Traditional detection techniques such as gas chromatography and infrared absorption provide high accuracy but are impractical for real-time applications due to high costs and operational complexity. Recent advancements in electronic nose systems and deep learning models have improved gas detection. However, challenges such as real-time inference and deployment on edge devices remain. This work introduces MENGLAN, a lightweight and efficient analyzer that addresses these challenges.

The electronic nose mimics the human olfactory system. It utilizes highly sensitive sensors to detect and identify odors and volatile organic compounds. Through specific algorithms, it achieves odor classification and quantitative analysis. The electronic nose is widely applied in various fields, including environmental monitoring \cite{domenech2023, yu2023}, food safety \cite{lu2022, huang2023, maimunah2020}, and medical diagnostics \cite{coronelteixeira2017, geffen2016, sun2017}, providing effective technical solutions for odor analysis.

In recent years, deep learning has been extensively used in electronic nose systems for gas concentration prediction in gas mixtures. Wu Jilong et al. \cite{wu2024}  combined RESHA and ALW, integrating the Hybrid Attention (HA) model to reweight features. Gan Wenchao et al. \cite{gan2025} integrated the Variational Mode Decomposition (VMD) method with the Long Short-Term Time-Series Network-Attention (LSTNet-Attention) model, enabling high-precision prediction of carbon monoxide and ethylene concentrations. Zeng Liwen et al. \cite{zeng2023} developed a Dual-Channel Temporal Convolutional Network (Dual-Channel TCN) based on Temporal Convolutional Networks (TCN), accurately determining the concentrations of multiple gas mixtures. Zhuo Junwei et al. \cite{zuo2025} designed the Parametric Rectified Linear Unit (PReLU)-based Multi-Head Attention Temporal Convolutional Network (PMH-TCN) model, effectively predicting gas concentrations. Xiong Lijian et al. \cite{xiong2023} introduced the combination of Gramian Angular Field (GASF/GADF) and Convolutional Neural Networks (CNN) into electronic nose data processing, classifying odor intensities. Li Juan et al. \cite{li2024} proposed a local dynamic neural network model, leveraging a pre-trained autoencoder network to achieve concentration prediction for multiple gases.

Existing gas concentration prediction models are accurate but suffer from slow inference and high computational demands, hindering real-time edge device deployment. Rapid gas concentration prediction is vital for chemical production safety, yet sensing device miniaturization limits computational power. Our proposed lightweight, high-sensitivity, multi-scale enhanced nonlinear gas analyzer addresses these issues. Its dual-stream architecture extracts local and global features, the multi-head attention mechanism processes key information in parallel, and the ReLU activation function improves the model's ability to capture nonlinear patterns in e-nose data. These improvements enable high-precision gas concentration prediction with reduced inference time, facilitating real-time ethylene monitoring and seamless deployment on compact sensing devices.

This study provides the following contributions:
\begin{itemize}
    \item We adopted a specialized information processing mode, minimizing raw data preprocessing to closely mimic real - world applications. To address the low signal - to - noise ratio, we designed the following modules:
    \item A dual - stream structure is employed to focus on special local features like mutations and missing data. In the local receptive field convolution feature stream, a new local receptive field information stream is introduced to identify feature patterns. By integrating global and local feature convolutions into one module, we fuse features at different scales, enabling the model to perceive and recognize global characteristics and achieve high - precision measurements of gas mixtures.
    \item We propose the Hybrid Multi - Head Attention Mechanism (HAMH) to identify self - attention weights across different receptive fields, facilitating the fusion of feature information at various scales.
    \item We developed the Feature Reconstruction Model (FRM). This module reconstructs the features of local receptive field data and preserves gradient information during model iteration through skip connections, focusing on skipped and low - information segments.
\end{itemize}

\section{DATASET DESCRIPTION AND PROCESSING}
\label{sec:2}
\subsection{Data Description}

To evaluate the performance of our proposed ethylene concentration prediction model, we conducted experiments using the public UCI dataset provided by Jordi Fonollosa \textit{et al.} \cite{fonollosa2015}. The dataset contains two types of gas mixtures: ethylene and carbon monoxide in air, as well as ethylene and methane in air. An array of 16 sensors was used to continuously collect signals at a frequency of 100 Hz for 12 hours, yielding over 1 million samples.

The measurement system consists of a data acquisition platform, a power control module, and a gas delivery system. The data acquisition platform's measurement chamber is equipped with four types of sensors, namely TGS - 2600, TGS - 2602, TGS - 2610, and TGS - 2620. During the experiment, the power control module maintained a constant voltage of 5 V for the sensors. The gas delivery system injected gases into the measurement chamber at a flow rate of 300 ml/min to capture the time response of different gas concentrations.

\begin{figure*}[htbp]
    \centering
    \includegraphics[width=0.6\textwidth]{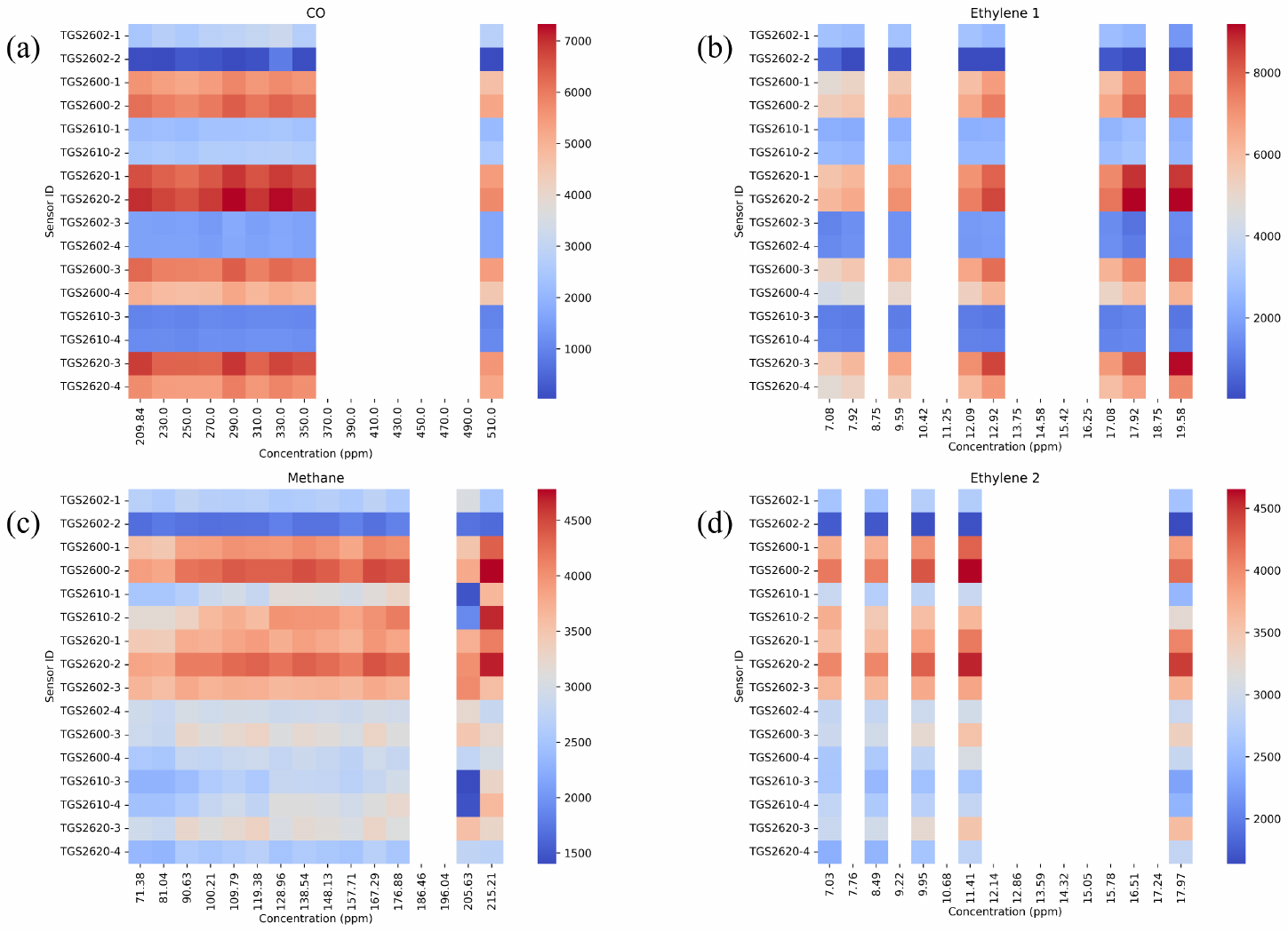} 
    \caption{(a) Thermodynamic distribution map of CO,
(b) Thermodynamic distribution map of ethylene in mixed gas 1,
(c) Thermodynamic distribution map of methane,
(d) Thermodynamic distribution map of ethylene in mixed gas 2.}
    \label{fig:heatmap}
\end{figure*}

Take the preparation of methane - ethylene mixtures as an example. Dry air, methane, and ethylene were fed into a pressurized cylinder controller (MFC) system through separate branches for mixing. The mixed gas was then injected into a 60 - ml measurement chamber. The sensors generated signals upon contact with the gas, which were transmitted to a PC and varied with gas concentration. Subsequently, the gas mixture was vented from the measurement chamber and collected by the exhaust system. The experiment recorded gas samples at 100 Hz for 12 hours, resulting in approximately 1 million samples. The heatmap distribution of the dataset is shown in Figure~\ref{fig:heatmap}.

\section{Methodology}
\label{sec:methodology}

\subsection{Data Processing}
The initial dataset includes 38 different concentration levels, comprising a total of 282 samples. To assess the predictive capability of the model, we retained the original data as much as possible, resulting in a dataset containing 282 samples. The dataset was then standardized using the following equation:

MENGLAN consists of three key components: Dual-Stream Structure for extracting local and global features, HMHA for combining spatial and temporal attention to optimize feature importance, and FRM to enhance feature representation and gradient flow.

\subsection{Dual-Stream Structure}
\begin{figure*}[htbp]
    \centering
    \includegraphics[width=0.8\textwidth]{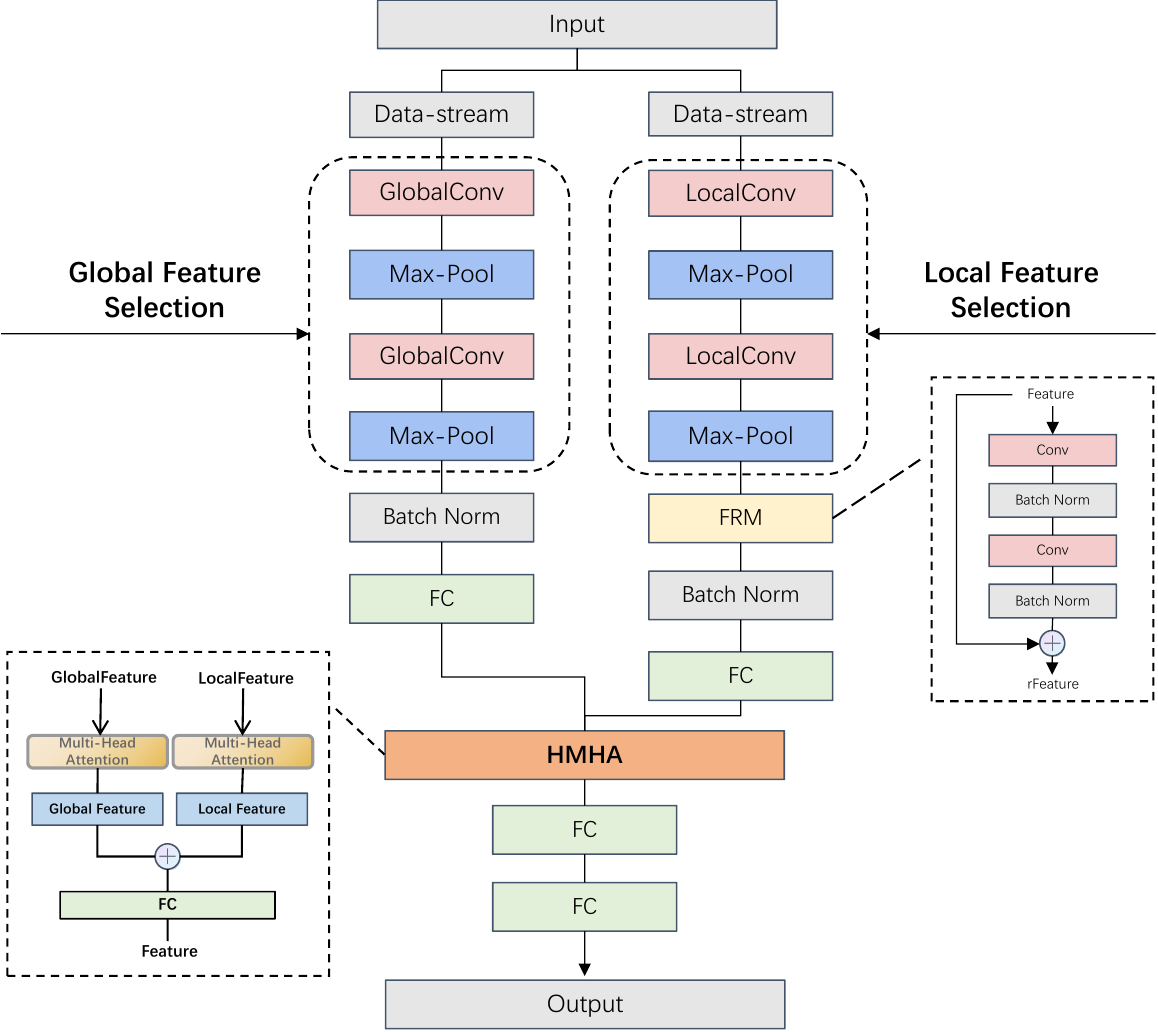} 
    \caption{MENGLAN integrates the FRM module for feature fusion activation and the HMHA mechanism for multi-scale convolutional feature fusion.}
    \label{fig:methodology}
\end{figure*}

We adopt a dual-stream structure to process global and local features through multiple convolutions and max-pooling operations in Figure~\ref{fig:methodology}. This design focuses on local feature irregularities while ensuring the model's global feature recognition and perception capabilities.Additionally, we employ the original ReLU as the activation function instead of other parametric ones (e.g., PReLU).

\subsubsection{Convolution}
In the dual-stream structure, we employ global and local convolutions for different feature processing requirements.

Global convolution applies the same kernel parameters across the entire feature dimension, reducing parameters through weight sharing while effectively extracting global patterns. This enables the network to detect similar structures across different spatial positions.

Local convolution (non-shared convolution) uses region-specific kernel parameters, allowing distinct feature representations for different areas. Despite increased parameter complexity, local convolution enhances the model's ability to capture special local features such as abrupt changes and missing data, improving generalization capability.

\subsubsection{Max-Pooling}
Max-pooling reduces feature map spatial dimensions by selecting the maximum value within sliding windows. This operation retains significant features while reducing computational complexity and improving network efficiency.

\subsection{Hybrid Multi-Head Attention (HMHA)}
To identify self-attention weights across different channels, we propose a Hybrid Multi-Head Attention mechanism (HMHA) in Figure~\ref{fig:methodology}, which integrates feature information from various receptive fields and scales. HMHA effectively enhances the model's representational power and feature integration capabilities by combining multi-head attention with local non-shared convolution operations. This approach allows for efficient extraction of both global contextual information and local position-specific patterns, thereby improving the model's ability to capture complex feature relationships.
\subsubsection{Multi-Head Attention}
Multi-head attention enhances the model's representational and parallelism capabilities. It captures various relationships within the input sequence through multiple parallel attention heads.
First, queries, keys, and values are projected into different subspaces via linear transformations. Then, each head computes attention outputs using scaled dot-product attention. The outputs from all heads are concatenated and projected into the final multi-head attention output.
\begin{equation}
\text{Output} = \text{Concat}(\text{Head}_1, \text{Head}_2, \dots, \text{Head}_h)W^O
\end{equation}
\begin{equation}
\text{Head}_i = \text{Attention}(QW_i^Q, KW_i^K, VW_i^V)
\end{equation}
where $W_i^Q$, $W_i^K$, and $W_i^V$ are the projection matrices, and $W^O$ is the output weight matrix.

\subsubsection{Local Non-Shared Convolution Operation}
In our HMHA mechanism, we employ local non-shared convolutions to capture position-specific patterns. Unlike traditional shared convolutions that use the same kernel across all spatial locations, non-shared convolutions apply different kernels to different spatial regions, enabling location-specific feature extraction.

For an input feature map $X \in \mathbb{R}^{H \times W \times C}$, the non-shared convolution operation is defined as:
\begin{equation}
Y_{i,j} = \sum_{u=-k}^{k} \sum_{v=-k}^{k} X_{i+u,j+v} \odot K_{i,j,u,v}
\end{equation}
where $Y_{i,j}$ is the output at spatial position $(i,j)$, $K_{i,j,u,v}$ is the position-specific kernel for location $(i,j)$ with offset $(u,v)$, and $\odot$ denotes element-wise multiplication. This approach enables the model to learn spatially-adaptive filters that are particularly effective for capturing local variations in the input data.

\subsection{Feature Reactivation Method}
The Feature Reactivation Module (FRM) improves feature utilization and network efficiency, especially in lightweight designs, by using skip connections and local data reconstruction. The process is as follows:
\begin{enumerate}
    \item Apply the first convolution with kernel $K_1$ to the input features $F$ to obtain $O_1$:
    \begin{equation}
    O_1 = F * K_1
    \end{equation}
    
    \item Normalize $O_1$ using the first batch normalization layer to obtain $BN_1$. During training, we use batch statistics:
    \begin{equation}
    BN_1^{(train)} = \frac{O_1 - \mu_1^{(batch)}}{\sqrt{(\sigma_1^{(batch)})^2 + \epsilon}} \cdot \gamma_1 + \beta_1
    \end{equation}
    During inference, we use running statistics:
    \begin{equation}
    BN_1^{(test)} = \frac{O_1 - \mu_1^{(running)}}{\sqrt{(\sigma_1^{(running)})^2 + \epsilon}} \cdot \gamma_1 + \beta_1
    \end{equation}
    where $\mu_1^{(batch)}$ and $(\sigma_1^{(batch)})^2$ are the batch-wise mean and variance during training, $\mu_1^{(running)}$ and $(\sigma_1^{(running)})^2$ are the running statistics used during inference, $\epsilon$ is a small constant, and $\gamma_1$, $\beta_1$ are learnable parameters for the first BN layer.
    
    \item Apply the second convolution to $BN_1$ to get $O_2$:
    \begin{equation}
    O_2 = BN_1 * K_2
    \end{equation}
    
    \item Normalize $O_2$ using the second batch normalization layer to obtain $BN_2$. During training:
    \begin{equation}
    BN_2^{(train)} = \frac{O_2 - \mu_2^{(batch)}}{\sqrt{(\sigma_2^{(batch)})^2 + \epsilon}} \cdot \gamma_2 + \beta_2
    \end{equation}
    During inference:
    \begin{equation}
    BN_2^{(test)} = \frac{O_2 - \mu_2^{(running)}}{\sqrt{(\sigma_2^{(running)})^2 + \epsilon}} \cdot \gamma_2 + \beta_2
    \end{equation}
    where $\mu_2^{(batch)}$, $(\sigma_2^{(batch)})^2$, $\mu_2^{(running)}$, $(\sigma_2^{(running)})^2$ are independent statistics for the second BN layer, and $\gamma_2$, $\beta_2$ are its corresponding learnable parameters.
    
    \item Combine $BN_2$ with the input features $F$ through a residual connection to produce the reactivated features $rFeature$:
    \begin{equation}
    rFeature = BN_2 + F
    \end{equation}
    
    \item The reactivated features $rFeature$ are subsequently processed through the HMHA mechanism and fed into the final prediction layers. The relationship between $rFeature$ and the model prediction $\hat{y}$ is expressed as:
    \begin{equation}
    \hat{y} = \text{Decoder}(\text{HMHA}(rFeature))
    \end{equation}
    where $\text{HMHA}(\cdot)$ represents the hybrid multi-head attention operation that processes the reactivated features, and $\text{Decoder}(\cdot)$ denotes the final prediction layers that map the attention-enhanced features to the target output space.
\end{enumerate}

\begin{table*}[h]
    \centering
    \caption{Model performance for ethylene in methane/ethylene and carbon monoxide/ethylene mixtures.}
    \label{tab:2}
    \begin{tabular}{@{}lrrrrccc@{}}
        \toprule[1.2pt]
        \textbf{Model} & \textbf{Parameters} & \textbf{RMSE} & \textbf{MSE} & \textbf{MAE} & \textbf{R\textsuperscript{2}} & \textbf{Total Inference} & \textbf{Average Inference} \\
        \midrule[0.9pt]
        \multicolumn{8}{@{}l}{\textit{Methane + Ethylene Mixture}} \\
        MENGLAN  & 8.93 MB  & 1.01  & 1.02  & 0.221 & 0.966 & 44.44\,s & 0.00005\,s \\
        MENGLAN  & 21.83 MB & 0.988 & 0.996 & 0.212 & 0.967 & 44.56\,s & 0.00005\,s \\
        MENGLAN  & 71.63 MB & 0.998 & 0.977 & 0.203 & 0.968 & 43.05\,s & 0.00005\,s \\
        2-ANN    & 17.02 MB & 1.475 & 2.175 & 0.605 & 0.928 & 44.74\,s & 0.00005\,s \\
        ANN      & 13.55 MB & 1.193 & 1.423 & 0.343 & 0.953 & 37.85\,s & 0.00005\,s \\
        CNN      & 4.31 MB  & 1.048 & 1.099 & 0.247 & 0.963 & 40.00\,s & 0.00005\,s \\
        CNN-RES  & 14.31 MB & 1.030 & 1.062 & 0.240 & 0.965 & 47.65\,s & 0.00006\,s \\
        
        \addlinespace[0.5em]
        \multicolumn{8}{@{}l}{\textit{Carbon Monoxide + Ethylene Mixture}} \\
        MENGLAN  & 8.93 MB  & 0.790 & 0.625 & 0.161 & 0.980 & 53.67\,s & 0.00006\,s \\
        MENGLAN  & 21.83 MB & 0.778 & 0.606 & 0.144 & 0.980 & 41.37\,s & 0.00005\,s \\
        MENGLAN  & 71.63 MB & 0.772 & 0.596 & 0.143 & 0.981 & 40.86\,s & 0.00005\,s \\
        2-ANN    & 17.02 MB & 1.548 & 2.398 & 0.710 & 0.924 & 43.71\,s & 0.00005\,s \\
        ANN      & 13.55 MB & 0.815 & 0.664 & 0.172 & 0.978 & 39.12\,s & 0.00005\,s \\
        CNN      & 4.31 MB  & 1.097 & 1.20  & 0.318 & 0.961 & 37.32\,s & 0.00004\,s \\
        CNN-RES  & 14.31 MB & 0.799 & 0.639 & 0.158 & 0.979 & 35.92\,s & 0.00004\,s \\
        \bottomrule[1.2pt]
    \end{tabular}
\end{table*}

\section{Result and Discussion}
\label{sec:4}
In this study, to ensure applicability and real-time performance in industrial scenarios, raw data was not downsampled, and missing information was retained. This approach ensures feasibility in practical use, as it is difficult to distinguish between missing data and systematic errors from measurement instruments. The method's effectiveness and accuracy were validated across two datasets and four prediction targets. Additionally, comparisons with four lightweight algorithms and models of varying parameter sizes demonstrated the superiority of the proposed method. Ablation experiments further illustrated the role and effectiveness of FRM and HMHA. Moreover, the performance of the model under different activation functions was analyzed. Consistent with common practice, we used MSE as the loss function during model updates, defined as Equation~(\ref{eq:mse}).
  \begin{equation}
\text{MSE} = \frac{1}{N} \sum_{t=1}^{N} (y_t - \hat{y}_t)^2
\label{eq:mse}
\end{equation}

Here, $y_t$ represents the true value of the target gas concentration, $\hat{y}_t$ denotes the predicted value, and $\bar{y}_t$ indicates the average value of the target gas.

\subsection{Evaluation Indicators}
To comprehensively evaluate model performance, several indicators were used. Root Mean Square Error (RMSE) measures the difference between predicted and true values; $R^2$ quantifies the model's fit to the dataset. Lower RMSE and SMAPE values indicate better fit, while $R^2$ values closer to 1 signify stronger fitting capability. Additionally, total and average inference time during the reasoning phase was considered to assess the model's deployment efficiency under lightweight conditions. These are defined as follows:

\begin{equation}
\text{RMSE} = \sqrt{\frac{1}{N} \sum_{t=1}^{N} (y_t - \hat{y}_t)^2}
\end{equation}

\begin{equation}
R^2 = 1 - \frac{\sum_{t=1}^{N} (y_t - \hat{y}_t)^2}{\sum_{t=1}^{N} (y_t - \bar{y}_t)^2}
\end{equation}

\subsection{Basic Information About the Experiment}
All experiments were conducted using an NVIDIA Tesla V100 GPU. The datasets were divided into training, validation, and testing sets in a 6:2:2 ratio. An early stopping strategy with a patience of 15 epochs was employed to supervise the learning process. Python version 3.8.5 and PyTorch version 2.0.0 were used. The network was trained for a maximum of 200 epochs with a batch size of 512. We used the Adam optimizer with a learning rate of 0.01 and decay rate of 0.08. The multi-head attention mechanism employed 8 attention heads, and ReLU was used as the activation function. A dropout rate of 0.2 was applied to prevent overfitting, with default values used for all other unspecified parameters.
\begin{table*}[htbp]
    \centering
    \caption{Prediction performance of different target gases in CO/ethylene and methane/ethylene mixtures}
    \label{tab:3}
    \begin{tabular}{@{}llrrrrcr@{}}
        \toprule[1.2pt]
        \textbf{TARGET} & \textbf{Model List} & \textbf{RMSE} & \textbf{MSE} & \textbf{MAE} & \textbf{R\textsuperscript{2}} & \textbf{Average Inference} \\
        \midrule[1.2pt]
        \multicolumn{7}{@{}l}{\textit{CO + Ethylene Mixture}} \\
        CO          & w/o FRM          & 23.144 & 535.652 & 4.639 & 0.979 & 0.00006\,s \\
        CO          & w/o HMHA         & 23.049 & 531.272 & 4.769 & 0.979 & 0.00004\,s \\
        CO          & w/o FRM \& HMHA  & 23.391 & 547.147 & 5.788 & 0.978 & 0.00004\,s \\
        C\textsubscript{2}H\textsubscript{4} & w/o FRM          & 0.876  & 0.769   & 0.221 & 0.975 & 0.00004\,s \\
        C\textsubscript{2}H\textsubscript{4} & w/o HMHA         & 0.931  & 0.867   & 0.237 & 0.972 & 0.00005\,s \\
        C\textsubscript{2}H\textsubscript{4} & w/o FRM \& HMHA  & 0.972  & 0.945   & 0.244 & 0.970 & 0.00005\,s \\
        
        \addlinespace[0.5em]
        \multicolumn{7}{@{}l}{\textit{Methane + Ethylene Mixture}} \\
        CO          & w/o FRM          & 16.874 & 284.751 & 4.498 & 0.951 & 0.00004\,s \\
        CO          & w/o HMHA         & 16.770 & 281.264 & 3.998 & 0.952 & 0.00004\,s \\
        CO          & w/o FRM \& HMHA  & 16.935 & 286.827 & 4.095 & 0.951 & 0.00004\,s \\
        C\textsubscript{2}H\textsubscript{4} & w/o FRM          & 1.024  & 1.048   & 0.258 & 0.965 & 0.00006\,s \\
        C\textsubscript{2}H\textsubscript{4} & w/o HMHA         & 1.274  & 1.623   & 0.360 & 0.946 & 0.00006\,s \\
        C\textsubscript{2}H\textsubscript{4} & w/o FRM \& HMHA  & 1.276  & 1.629   & 0.372 & 0.946 & 0.00005\,s \\
        \bottomrule[1.2pt]
    \end{tabular}
\end{table*}

\subsection{Experimental Results}
\subsubsection{Comparison with Other Methods}
This work compares the prediction performance for ethylene concentration across two datasets with ANN\cite{ren_gas_2022,liu_anti-interference_2024}, CNN\cite{chen2024dea}, 2-ANN\cite{zeng_mixed_2023}, CNN-RES, and MENGLAN models of different parameter sizes. Tables~\ref{tab:2} present the results.

\begin{figure*}[!htbp]
    \centering
    \includegraphics[width=1\textwidth]{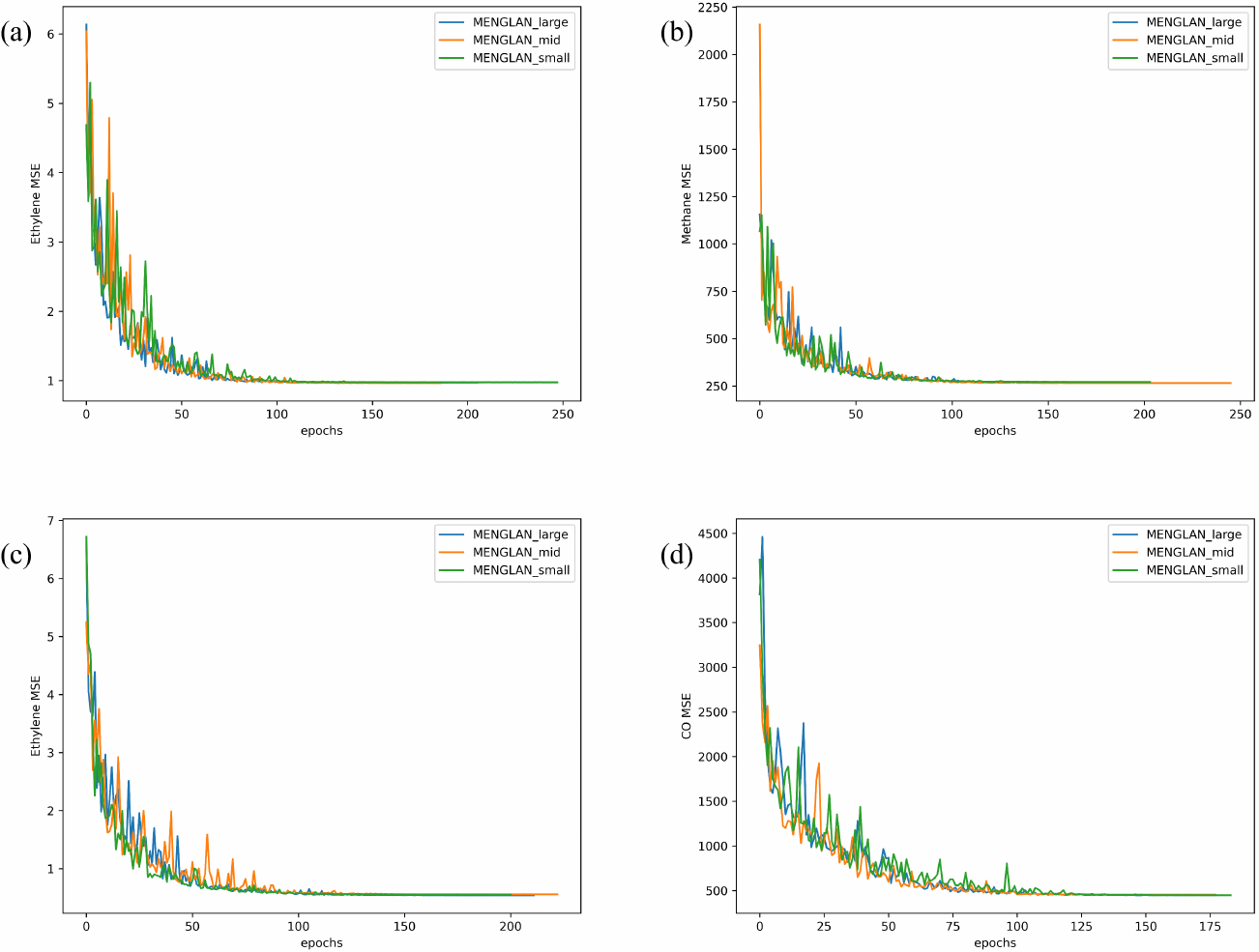} 
    \caption{Error comparison of models with different parameter amounts during the training process:
(a) MSE for ethylene,
(b) MSE for methane,
(c) MSE for ethylene,
(d) MSE for CO.}
    \label{fig:error_analysis}
\end{figure*}

From Tables~\ref{tab:2} , MENGLAN outperforms all models in RMSE, MSE, MAE, and $R^2$, showcasing its effectiveness and scalability. Figures~\ref{fig:error_analysis}  illustrate training error trends and performance comparisons for models with varying parameter sizes, respectively.

\begin{table*}[htbp]
    \centering
    \caption{Performance of predicting gas concentrations in the mixture using three different activation functions on the test set.}
    \label{tab:4}
    \begin{tabular}{@{}lrrrrccc@{}}
        \toprule[1.2pt]
       \textbf{Activation} & \textbf{TARGET} & {\textbf{RMSE}} & {\textbf{MSE}} & {\textbf{MAE}} & {\textbf{R\textsuperscript{2}}} & {\textbf{Total inference (\si{\second})}} \\
\midrule
tanh & C2H4 & 1.683 & 2.834 & 0.657 & 0.907 & 40.16 \\
     & CH4  & 28.301 & 800.964 & 11.254 & 0.864 & 49.49 \\
     & C2H4 & 2.279 & 5.189 & 1.050 & 0.836 & 41.71 \\
     & CO   & 69.335 & 4807.373 & 30.625 & 0.814 & 41.58 \\
\midrule
sigmoid & C2H4 & 1.788 & 3.196 & 0.680 & 0.895 & 41.29 \\
        & CH4  & 40.040 & 1603.257 & 19.730 & 0.727 & 40.70 \\
        & C2H4 & 4.491 & 20.171 & 3.607 & 0.3627 & 41.42 \\
        & CO   & 160.736 & 25836.163 & 148.666 & \(1.19 \times 10^{-7}\)& 41.50 \\
\midrule
softsign & C2H4 & 1.451 & 2.107 & 0.470 & 0.931 & 41.61 \\
         & CH4  & 24.000 & 576.009 & 8.559 & 0.902 & 41.11 \\
         & C2H4 & 1.261 & 1.589 & 0.388 & 0.950 & 41.79 \\
         & CO   & 43.771 & 1915.899 & 15.913 & 0.926 & 42.07 \\
\bottomrule
    \end{tabular}
\end{table*}
\subsubsection{Ablation Experiment}

In this section, we further ablate different modules to validate their computational contributions, keeping all other experimental settings consistent.Table~\ref{tab:3} illustrate the prediction performance of different target gases in two gas mixtures, respectively.

Table~\ref{tab:3} indicate that the FRM and HMHA methods significantly enhance computational efficiency. Excluding the FRM module reduces the model's ability to extract local features, making it challenging to capture abrupt signals from raw data, thereby degrading performance. Similarly, excluding the HMHA module diminishes the model's capacity for weighted integration of global and local features, leading to performance loss.

\subsubsection{Comparison of Different Activation Functions}

Considering the significant impact of activation functions on model performance, we explore the effects of replacing the ReLU function in the model with Sigmoid, Softsign, and Tanh functions. Experiments were conducted with consistent parameter settings, and the results are presented in Table~\ref{tab:4}.
The experimental conclusions can be drawn from a comparison of Tables~From \ref{tab:2} and \ref{tab:4}. It can be observed that ReLU consistently outperforms other activation functions under various conditions. Therefore, when ReLU is chosen as the activation function, the MENGLAN model consistently achieves satisfactory results across different scenarios.

\section{Conclusion}
\label{sec:conclusion}

This study successfully developed MENGLAN, an instrument for real-time ethylene concentration prediction in chemical production gas mixtures. Its dual-stream structure, HMHA mechanism, and FRM model minimized preprocessing steps and achieved high accuracy and fast inference speeds. Experimental results showed reduced prediction errors and improved dataset fitting. MENGLAN offers a feasible solution for edge device deployment, enhancing production safety and efficiency. Future work could explore scalability and further optimization.

\printbibliography

@article{domenech2023,
  author = {Domènech Gil Guillem and Duc Nguyen Thanh and Wikner J Jacob and Eriksson Jens and Påledal Sören Nilsson and Puglisi Donatella and Bastviken David},
  title = {Electronic Nose for Improved Environmental Methane Monitoring},
  journal = {Environmental Science \& Technology},
  volume = {57},
  number = {12},
  pages = {4567--4576},
  year = {2023}
}

@article{yu2023,
  author = {Yu Zhichong and Tian Xu and Gao Yichun and Yuan Xuehong and Xu Zhenming and Zhang Lingen},
  title = {Monitoring the Resources and Environmental Impacts from the Precise Disassembly of E-Waste in China},
  journal = {Environmental Science \& Technology},
  volume = {57},
  number = {22},
  pages = {8900--8909},
  year = {2023}
}

@article{lu2022,
  author = {Lu Lin and Hu Zhanqiang and Hu Xianqiao and Li Dan and Tian Shiyi},
  title = {Electronic tongue and electronic nose for food quality and safety},
  journal = {Food Research International},
  volume = {154},
  number = {PB},
  pages = {112214},
  year = {2022}
}

@article{huang2023,
  author = {Huang Gui-Li and Liu Tian-Tian and Mao Xiao-Mei and Quan Xin-Yao and Sui Si-Yao and Ma Jia-Jia and Wang Yu-Ning},
  title = {Insights into the volatile flavor and quality profiles of loquat (Eriobotrya japonica Lindl.) during shelf-life via HS-GC-IMS, E-nose, and E-tongue},
  journal = {Food Chemistry: X},
  volume = {100886},
  pages = {1--11},
  year = {2023}
}

@article{maimunah2020,
  author = {Maimunah Mohd Ali and Norhashila Hashim and Samsuzana Abd Aziz and Ola Lasekan},
  title = {Principles and recent advances in electronic nose for quality inspection of agricultural and food products},
  journal = {Trends in Food Science \& Technology},
  volume = {prepublish},
  number = {1},
  pages = {1--10},
  year = {2020}
}

@article{sun2017,
  author = {Sun Hao and Tian Fengchun and Liang Zhifang and Sun Tong and Yu Bin and Yang Simon X and Liu Xiangmin},
  title = {Sensor Array Optimization of Electronic Nose for Detection of Bacteria in Wound Infection},
  journal = {IEEE Transactions on Industrial Electronics},
  volume = {9},
  number = {7350},
  pages = {1--9},
  year = {2017}
}

@article{coronelteixeira2017,
  author = {Rosarito Coronel Teixeira and Mabel Rodríguez and Nilda Jiménez de Romero and Marcel Bruins and Roscio Gómez and Jan Bart Yntema and Cecile Magis-Escurra},
  title = {The potential of a portable, point-of-care electronic nose to diagnose tuberculosis},
  journal = {Journal of Infection},
  volume = {5},
  number = {441},
  pages = {447},
  year = {2017}
}

@article{geffen2016,
  author = {van Geffen Wouter H. and Kerstjens Huib},
  title = {Diagnosing viral and bacterial respiratory infections in acute COPD exacerbations by electronic nose},
  journal = {EUROPEAN RESPIRATORY JOURNAL},
  volume = {s60},
  year = {2016}
}

@article{wu2024,
  author = {Jilong Wu and Wenlong Zhao and Fan Wu and Jia Yan and Peter Feng and Hao Cui and Xiaoyan Peng},
  title = {A mixed gas concentration regression prediction method based on RESHA-ALW},
  journal = {Sensors and Actuators: B. Chemical},
  volume = {136222},
  pages = {1--10},
  year = {2024}
}

@article{gan2025,
  author = {Wenchao Gan and Ruilong Ma and Wenlong Zhao and Xiaoyan Peng and Hao Cui and Jia Yan and Jin Chu},
  title = {A VMD-LSTNet-Attention model for concentration prediction of mixed gases},
  journal = {Sensors and Actuators: B. Chemical},
  volume = {136641},
  pages = {1--12},
  year = {2025}
}

@article{zeng2023,
  author = {Zeng Liwen and Xu Yang and Ni Sen and Xu Min and Jia Pengfei},
  title = {A mixed gas concentration regression prediction method for electronic nose based on two-channel TCN},
  journal = {Sensors and Actuators: B. Chemical},
  year = {2023}
}

@article{zuo2025,
  author = {Junwei Zhuo and Xingyu Chen and Huisheng Zhang and Xue Wang and Pengcheng Wu and Jiaxin Yue and Jin Chu},
  title = {Rapid and high-accuracy concentration prediction of gas mixtures based on PMH-TCN},
  journal = {Measurement},
  volume = {PC},
  number = {116003},
  pages = {1--8},
  year = {2025}
}

@article{xiong2023,
  author = {Xiong Lijian and He Meng and Hu Can and Hou Yuxin and Han Shaoyun and Tang Xiuying},
  title = {Image presentation and effective classification of odor intensity levels using multi-channel electronic nose technology combined with GASF and CNN},
  journal = {Sensors and Actuators: B. Chemical},
  year = {2023}
}

@article{li2024,
  author = {Li Juan and Ma Yilun and Duan Zaihua and Zhang Yajie and Duan Xiaohui and Liu Bohao and Tai Huiling},
  title = {Local dynamic neural network for quantitative analysis of mixed gases},
  journal = {Sensors and Actuators: B. Chemical},
  volume = {135230},
  pages = {1--9},
  year = {2024}
}

@article{fonollosa2015,
  author = {Jordi Fonollosa and Sadique Sheik and Ramón Huerta and Santiago Marco},
  title = {Reservoir computing compensates slow response of chemosensor arrays exposed to fast varying gas concentrations in continuous monitoring},
  journal = {Sensors \& Actuators: B. Chemical},
  volume = {618},
  pages = {629},
  year = {2015}
}

@article{chen2024dea,
  title   = {DEA-Net: Single Image Dehazing Based on Detail-Enhanced Convolution and Content-Guided Attention},
  author  = {Chen, Zixuan and He, Zewei and Lu, Zhe-Ming},
  journal = {IEEE Transactions on Image Processing},
  volume  = {33},
  pages   = {1002--1015},
  year    = {2024},
  publisher= {IEEE},
  
}

@article{ren_gas_2022,
  author = {Ren, Wenjie and Zhao, Changhui and Niu, Gaoqiang and Zhuang, Yi and Wang, Fei},
  title = {Gas Sensor Array with Pattern Recognition Algorithms for Highly Sensitive and Selective Discrimination of Trimethylamine},
  journal = {Advanced Intelligent Systems},
  year = {2022},
  volume = {4},
  number = {2200169},
  pages = {2200169 (1-10)},
  
  
}

@article{liu_anti-interference_2024,
  author = {Liu, Yupeng and Yang, Zhuang and Huang, Long and Zeng, Wen and Zhou, Qu},
  title = {Anti-interference detection of mixed {NOX} via {In2O3}-based sensor array combining with neural network model at room temperature},
  journal = {Journal of Hazardous Materials},
  year = {2024},
  volume = {463},
  pages = {132857},
  
}

@article{zeng_mixed_2023,
  author = {Zeng, Liwen and Xu, Yang and Ni, Sen and Xu, Min and Jia, Pengfei},
  title = {A mixed gas concentration regression prediction method for electronic nose based on two-channel {TCN}},
  journal = {Sensors and Actuators: B. Chemical},
  year = {2023},
  volume = {382},
  pages = {133528},
  
}

\end{document}